\newcommand{\RR}{\textsf{ReveR}}
\newcommand{\RN}{\textsf{4104}}

\documentclass{article}
\usepackage{amssymb,multicol,url}
\usepackage{graphicx}

\begin{document}

\title{\bf\RR: Software Simulator of Reversible Processor with Stack}
\author{Alexander Yu. Vlasov%
\thanks{Electronic mail: {\tt alexander.vlasov@pobox.spbu.ru}}%
}

\maketitle

\begin{abstract}
A software model of a reversible processor {\sf ReveR} with the stack is discussed 
in this paper. An architecture, the minimal set of elementary reversible
operations together with an implementation of the basic control flow structures 
and procedures calls using simple assembler language are described. 
\end{abstract}

\section{Introduction}

An origin of this work was the research of models of programmable quantum processors 
\cite{PQP},  but this presentation does not require any knowledge about quantum 
algorithms. A program simulator of reversible processor \RR\ is discussed. It may be 
useful for the research of the reversible computation even without an application 
to the quantum information science. Irreversible instructions may be implemented 
using the idea of history tape \cite{BenRev}.

The simulator uses minimal set of operations briefly represented below without 
a discussion and comparison with many well known modern models of reversible 
processors \cite{Pen,Pisa,TGA}. The main purposes of the current version is 
the consideration of the control flow in a reversible processor, e.g. unconditional 
and conditional jumps, loops and calls of procedures using the stack for a return 
address. The reversible implementation of procedure calls and basic control flow 
structures with minimal set of basic instructions is demonstrated .

\section{Architecture}

\subsection{Registers and Basic Processing Principles}

The current model of the processor \RR\ has eight 32-bit registers:
{\tt CM} --- the current instruction (command), {\tt IP} --- the instruction pointer, 
{\tt DIP} --- the increment (delta) of the instruction pointer, {\tt SP} --- the
stack pointer, {\tt MP} --- the memory pointer, {\tt RA}, {\tt RB}, {\tt RC} --- 
registers {\em A,B,C}. 

A standard step formally uses two phases, an external and an internal. 
Between the steps the register {\tt CM} has zero value and {\tt IP} points
to the address of a current instruction. 
The external phase is fixed and changes only the registers {\tt CM} and {\tt IP}
and the internal phase may change all, except these two registers.
\begin{verbatim} 
 CM := CM + MEM[IP];
  Internal(CM);
 CM := CM - MEM[IP]; IP := IP + DIP
\end{verbatim}
The internal phase is also quite simple. It is used an idea to choose a minimal set
of elementary reversible operations and to represent an instruction as the sequence
of such operations without any internal branching and changes of the {\tt CM}, {\tt IP}
registers. It guarantees the reversibility of any instruction and a relative simplicity
of a program flow. The processor \RR\ uses an array of ``nanoprograms''\footnote{Such 
an informal term is used here instead of more common ``microprogram,'' 
due to actuality of the quantum and reversible computations in nanoscale.},
i.e. sequences of indexes of the reversible instructions. 
\begin{verbatim}
  with Nanoprogram[CM] do
   for nIP := 0 to Length(RevCodes) - 1 do
    RevOps[RevCodes[nIP]].Perform
\end{verbatim}
Here is used an internal processor counter denoted as {\tt nIP}.
{\tt RevCodes} is an array of codes for given instruction, 
{\tt Nanoprogram[CM]} and {\tt RevOps} points to a reversible
operation for given code (see the table \ref{TOpCod} below with examples of the 
codes, {\tt n} and associated operations).
For any instruction is known an inverse one and so such step is reversible.
The inversion of a nanoprogram should perform inverse operations
in opposite order 
\begin{verbatim}
{...}  for nIP := Length(RevCodes) - 1 downto 0 do  {...}
\end{verbatim}

\subsection{Elementary Internal Operations}

For the simplest implementation it is enough to use sixteen elementary
instructions shown in the table \ref{TOpCod}. A short 
notation {\tt [MP]=MEM[MP]} for contents of memory with address {\tt MP} and 
{\tt [IP+1]=MEM[IP+1]} for a number {\tt NX} immediately after current instruction
is used in the table.
It is also used notation $A \leftrightarrow B$ for the exchange and
$\hookleftarrow$ for the left assignment in the reversible operations
like $A \hookleftarrow B - A$.
Yet another shortcut is the step-like function $G_C$ equal to unit for
any positive $C$ and zero otherwise. So the operation {\tt CNSUB}
may be written as 
\mbox{\tt A = (C > 0) ? B - A : -A}
in a C-like language.

\begin{table}[htb]
\begin{center}
\setlength{\tabcolsep}{3pt}
\begin{tabular}{|r|l|c|l||r|l|c|l|}
\hline
{\tt n} & Name & Description & Inverse &
{\tt n} & Name & Description & Inverse \\
\hline
0 & \tt NOP & No operation & \tt NOP &
8 & \tt SWP MEM & $A \leftrightarrow {\tt [MP]}$  & \tt SWP MEM \\ 
1 & \tt NOT RA & $A \hookleftarrow -1{-}A$ & \tt NOT RA & 
9 & \tt NSUB & $A \hookleftarrow B - A$  & \tt NSUB\\ 
2 & \tt NEG RA & $A \hookleftarrow - A $ & \tt NEG RA &
10 & \tt NX SUB & $A \hookleftarrow {\tt [IP{+}1]} - A$  & \tt NX SUB \\   
3 & \tt SWP RB & $A \leftrightarrow B$  & \tt SWP RB &  
11 & \tt NSUB IP& $A \hookleftarrow {\tt IP} - A$  & \tt NSUB IP\\ 
4 & \tt SWP RC & $A \leftrightarrow C$  & \tt SWP RC &
12 & \tt CNSUB  & $ A \hookleftarrow G_C B - A $  & \tt CNSUB \\ 
5 & \tt SWP DIP & $A \leftrightarrow {\tt DIP}$  & \tt SWP DIP &  
13 & \tt CNX SUB  & $ A \hookleftarrow G_C {\tt [IP{+}1]} - A $  & \tt CNX SUB \\
6 & \tt SWP SP & $A \leftrightarrow {\tt SP}$  & \tt SWP SP &  
14 & \tt CLRA & $0 \looparrowright A \looparrowright $ History & \tt UCLA \\
7 & \tt SWP MP & $A \leftrightarrow {\tt MP}$  & \tt SWP MP &  
15 & \tt UCLA & $0 \looparrowleft A \looparrowleft $ History & \tt CLRA \\
\hline
\end{tabular}
\end{center}
\caption{Elementary codes for \RR.}\label{TOpCod}
\end{table}

The operations with codes 0 -- 13 are self-inverse (involutions).
The exceptions from such a rule are {\tt CLRA} and {\tt UCLA}
necessary only for the simulation of irreversible operations with 
a ``history tape'' and mentioned for completeness. 
{\tt CLRA} ``pushes'' value of {\tt RA} to a ``trash'' and 
supplies {\tt RA} with ``new'' zero.
The inverse operation $\tt UCLA$ should be never used directly and may appears
only during the reverse running of the processor. In such a case
an application of $\tt UCLA$ to nonzero {\tt RA} may appears only due 
to some error and results an exception in the \RR. 

\subsection{Memory Access}

The direct access of the \RR\ to a memory is ensured due to the instruction {\tt SWP~MEM}
via an exchange of {\tt RA} and the memory with an address stored in {\tt MP}. In 
the current model data, addressed by {\tt MP}, instructions, addressed by {\tt IP}
and the stack share the same physical memory. After initialization of the registers,
of the \RR\, {\tt SP} points to the last address of the memory and all other registers 
are zero. An access to the stack is discussed further and uses simple nanoprograms 
with the temporary exchange {\tt MP} and {\tt SP}.

Yet another method to obtain some number is provided by the access to
the address immediately after {\tt IP} ({\tt NX = [IP+1]}) using operations 
such as {\tt NX SUB} and {\tt CNX SUB}. An operation for a direct exchange
of some register with {\tt [IP+1]} is not provided because formally it could
change program itself, but it would not be in agreement with some principles of
suggested architecture. The operation {\tt NSUB IP} also could provide a similar
functionality in a less direct way and may be considered as redundant.

\section{Instructions and Control Flow}

\subsection{Basic Instructions (``Nanoprograms'')}

\begin{table}[htb]
\begin{center}
\setlength{\tabcolsep}{2pt}
\noindent\begin{tabular}{|c|l|l|}
\hline
{\tt CM} & Name & Description \\ \hline
0 & \tt Nop & \tt NOP\\ \hline
1 & \tt Inc DIP & $\tt DIP \hookleftarrow DIP {+} 1$\\ \hline
2 & \tt Dec DIP & $\tt DIP \hookleftarrow DIP {-} 1$\\ \hline
3 & \tt Inc RA  & $A \hookleftarrow A + 1$\\ \hline
4 & \tt Dec RA  & $A \hookleftarrow A - 1$\\ \hline
5--10 & \tt Swap RA,\ldots & $A \leftrightarrow \ldots$\\ \hline
11 & \tt Add RA,RB &$A \hookleftarrow A + B$\\ \hline
12 & \tt Sub RA,RB &$A \hookleftarrow A - B$\\ \hline
13 & \tt Add RA,[IP+1] &$A {\hookleftarrow} A {+} [{\tt IP}{+}1]$\\ \hline
14 & \tt Sub RA,[IP+1] &$A {\hookleftarrow} A {-} [{\tt IP}{+}1]$\\ \hline
15 & \tt Add RA,IP &$A \hookleftarrow A + {\tt IP}$\\ \hline
16 & \tt Sub RA,IP &$A \hookleftarrow A - {\tt IP}$\\ \hline
\end{tabular}\,%
\renewcommand{\arraystretch}{1.092}%
\begin{tabular}{|c|l|l|}
\hline
{\tt CM} & Name & Description \\ \hline
17 & \tt Neg RA &$A \hookleftarrow - A $\\ \hline
18 & \tt Jmp dNx &$\tt DIP \hookleftarrow DIP + [IP{+}1]$\\ \hline
19 & \tt Subr    &$\tt DIP \leftrightarrow [SP]; 
  DIP \hookleftarrow DIP + 1;$\\
   &             &$ \tt SP \hookleftarrow SP - 1 $\\ \hline   
20 & \tt Ret     &$\tt SP \hookleftarrow SP + 1; DIP \hookleftarrow DIP - 1;$ \\            
   &             &$\tt DIP {\leftrightarrow} [SP]; DIP \hookleftarrow [IP{+}1]{-}DIP$ \\ \hline
21 & \tt CJmp dNx   & if $C > 0,~ \tt DIP \hookleftarrow DIP + [IP{+}1]$\\ \hline
22 & \tt CAdd RA,RB & if $C > 0,~ A \hookleftarrow A + B$ \\ \hline
23 & \tt Push RA  &$ A  \leftrightarrow \tt[SP];   SP \hookleftarrow SP - 1 $ \\ \hline
24 & \tt Pop RA  & ${\tt SP \hookleftarrow SP + 1; [SP]} \leftrightarrow A $  \\ \hline
25 & \tt Clear RA & $0 \looparrowright A \looparrowright $ History \\ \hline
\end{tabular}
\end{center}
\caption{Descriptions of basic instructions for \RR\RN.}\label{TNanoOp}
\end{table}

In the table \ref{TNanoOp} are provided the description of nanoprograms for the
basic set of instructions used in the current version of the processor \RR\RN. 
Simplest instructions such as {\tt Nop}, {\tt Swap RA,\ldots} are equivalent
to single internal operations, other nanoprograms may call from 
2 to 14 elementary operations. The codes of the operations are provided 
in the table \ref{TNanoCod}.

\begin{table}[htb]
\begin{center}
\setlength{\tabcolsep}{4pt}
\begin{tabular}{|c|l|l|}
\hline
{\tt CM} & Name & Internal codes \\
\hline
0 & \tt Nop & \tt NOP\\ \hline
1 & \tt Inc DIP &\tt SWP DIP; NOT RA; NEG RA; SWP DIP\\ \hline
2 & \tt Dec DIP &\tt SWP DIP; NEG RA; NOT RA; SWP DIP\\ \hline
3 & \tt Inc RA &\tt NOT RA; NEG RA\\ \hline
4 & \tt Dec RA &\tt NEG RA; NOT RA\\ \hline
5 & \tt Swap RA,RB &\tt SWP RB\\ \hline
6 & \tt Swap RA,RC &\tt SWP RC\\ \hline
7 & \tt Swap RA,DIP &\tt SWP DIP\\ \hline
8 & \tt Swap RA,SP &\tt SWP SP\\ \hline
9 & \tt Swap RA,MP &\tt SWP MP\\ \hline
10 & \tt Swap RA,[MP] &\tt SWP MEM\\ \hline
11 & \tt Add RA,RB &\tt NEG RA; NSUB\\ \hline
12 & \tt Sub RA,RB &\tt NSUB; NEG RA\\ \hline
13 & \tt Add RA,[IP+1] &\tt NEG RA; NX SUB\\ \hline
14 & \tt Sub RA,[IP+1] &\tt NX SUB; NEG RA\\ \hline
15 & \tt Add RA,IP &\tt NEG RA; NSUB IP\\ \hline
16 & \tt Sub RA,IP &\tt NSUB IP; NEG RA\\ \hline
17 & \tt Neg RA & \tt NEG RA\\ \hline
18 & \tt Jmp dNx &\tt SWP DIP; NEG RA; NX SUB; SWP DIP\\ \hline
19 & \tt Subr    &\tt SWP SP; SWP MP; SWP SP; \\
   &             &\tt SWP DIP; SWP MEM; NOT RA; NEG RA; SWP DIP;\\
   &             &\tt SWP SP; SWP MP; SWPSP; NEG RA; NOT RA; SWP SP\\ \hline   
20 & \tt Ret     &\tt SWP SP; NOT RA; NEG RA; SWP MP, SWP SP; SWP DIP;\\
   &             &\tt NEG RA; NOT RA; SWP MEM; SWP SP; SWP MP; SWPSP;\\ 
   &             &\tt NX SUB; SWP DIP \\ \hline
21 & \tt CJmp dNx   & \tt SWP DIP; NEG RA; CNX SUB; SWP DIP\\ \hline
22 & \tt CAdd RA,RB & \tt NEG RA; CNSUB\\ \hline
23 & \tt Push RA  & \tt SWP SP; SWP MP; SWP SP; SWP MEM; \\
   &              & \tt SWP SP; SWP MP; NEG RA; NOT RA; SWP SP\\ \hline
24 & \tt Pop RA   & \tt SWP SP; NOT RA; NEG RA; SWP MP; SWP SP;\\
   &              & \tt SWP MEM; SWP SP; SWP MP; SWP SP  \\ \hline
25 & \tt Clear RA & \tt CLRA \\
\hline
\end{tabular}
\end{center}
\caption{Codes (nanoprograms) for basic instructions of \RR\RN.}\label{TNanoCod}
\end{table}

Any program should start
with {\tt Inc DIP}, because the \RR\ is initialized with zero {\tt DIP}
register and without such instruction {\tt IP} would not increase 
after each step.   
Before instructions such as {\tt Add RA,[IP+1]} (with am immediate number in 
the next address) {\tt DIP} should be increased again to make double steps. 

The instruction 
{\tt Jmp dNx} uses an immediate number to increase {\tt DIP} and
make the unconditional jump. A destination address of such a jump also 
should contain {\tt Jmp dNx} with a negative increment to restore 
an initial value of {\tt DIP}. 

An example is provided below.
Here semicolons are used for comments, colons for labels (like @1, @2) and the
\# prefix for immediate values, e.g. \#@2-@1-1 is equal to the difference 
between addresses marked by labels @2 and @1 decreased by unit.
\begin{verbatim} 
Inc DIP   ; To start program, DIP = 1
:@1       ; Label before jump instruction
Jmp dNx   ; Jump instruction (set DIP = @2-@1)
#@2-@1-1  ; Immediate value, increment of DIP
Nop       ; To skip next instructions
Nop
:@2       ; Label to jump here
Jmp dNx   ; "Fake" jump instruction, set DIP = 2 
#@1-@2+2  ; Immediate negative value, decrement of DIP
Dec DIP   ; To set DIP = 1
\end{verbatim}

The application of the conditional jump {\tt CJmp dNx} is similar and may be
used for\\ ``\mbox{\tt if RC > 0 then \ldots}'' structures. 
\begin{verbatim} 
Inc DIP   ; To start program, DIP = 1
Inc RA    ; RA = 1
Swap RA,RC  ; RC = 1
Inc DIP   ; Set DIP = 2 to omit immediate value after jump
Nop       ; The gap is necessary due to DIP = 2 
:@1       ; Label before jump instruction
CJmp dNx  ; Conditional jump (if RC > 0 then DIP = @2-@1)
#@2-@1-2  ; Immediate value, increment of DIP
Nop       ; Skipped, because RC > 0
Nop
:@2       ; Label to jump here
Jmp dNx   ; "Fake" jump instruction, set DIP = 2 
#@1-@2+2  ; Immediate negative value, decrement of DIP
Dec DIP   ; To set DIP = 1
\end{verbatim}

Different control structures with loops also may be implemented with 
{\tt CJmp dNx} commands.
An example below shows a loop like: {\tt For RC := 0 to 9 do Inc RB}.

{\setlength{\columnseprule}{.4pt}
\begin{multicols}{4}
\begin{verbatim}
Inc DIP
; to start loop
Inc DIP
Nop
:@1
CJmp dNx
#@2-@1+2
Dec DIP
; Inc RB
Swap RA,RB
Inc RA
Swap RA,RB
; If RC + 1 > 9
Swap RA,RC
Inc RA
Inc DIP
Nop
Sub RA,[IP+1]
#9 ; count
Swap RA,RC
Nop
:@3 ; exit loop
CJmp dNx
#@4-@3-2
Swap RA,RC
Nop
Add RA,[IP+1]
#9 ; count
Dec DIP
Swap RA,RC
; continue
Inc DIP
Nop
:@2
CJmp dNx
#@1-@2-2
:@4
CJmp dNx
#@3-@4+2
Dec DIP
; end of loop 
Swap RA,RC
Dec RA
Swap RA,RC
\end{verbatim}
\end{multicols} 
}

The more difficult case is a procedure call, because it requires the tracing of
the return address if the same procedure may be called from different
locations. The stack should be used for the return positions to allow 
nesting of the calls. 

\subsection{Procedures and Reversible Stack}

Instruction {\tt Push RA} and {\tt Pop RA} represented in table \ref{TNanoOp}
look simple, yet each uses nine reversible internal codes due to necessity
of auxiliary swaps of {\tt MP}, {\tt SP} and {\tt RA}. The instruction {\tt Push RA}
exchanges {\tt RA} with content of memory with address stored in {\tt SP}
and decreases {\tt SP} on unit to point on previous address. The {\tt Pop RA}
is inverse of {\tt Push RA}.

It is suggested for proper work, that {\tt SP} always points to address with
zero value to ensure {\tt RA = 0} after {\tt Push RA}. So, it is necessary to keep 
zero value of {\tt RA} before {\tt Pop RA}.

So, if contents of {\tt RA} is not known, it should be used {\tt CLRA}
command and it is not purely reversible computations. It may be shown, 
that procedure calls with stack may be performed without {\tt CLRA}.

Simple program below demonstrates principle of such reversible calls.
{\setlength{\columnseprule}{.4pt}
\begin{multicols}{3}
\begin{verbatim}
; To start program
Inc DIP 
:@1 ; First call
Jmp dNx ; ->   to @3
#@3-@1-1
Jmp dNx ; <- from @4
#@4-@1
Dec DIP ; DIP = 1
:@2 ; Second call
Jmp dNx ; ->   to @3
#@3-@2-1
Jmp dNx ; <- from @4
#@4-@2  
Dec DIP ; DIP = 1
; ... etc.
:@3 ;Procedure label
Subr ; To push DIP
; ... do something
:@4 ;Return label 
Ret  ; Return
#@3-@4+2  
\end{verbatim}
\end{multicols}
}

Already mentioned {\tt Jmp dNx} is used here, but the {\tt Subr} instruction 
at very beginning of the procedure stores {\tt DIP} in the stack. The instruction
{\tt Ret} at end of the procedure restores value of {\tt DIP} from the stack and uses
the immediate value (@2-@3+2) to calculate the size of the jump to an instruction after
the procedure call. This instruction should be {\tt Jmp dNx} with an appropriate 
immediate negative value to set {\tt DIP = 2} (followed by {\tt Dec DIP} to set 
{\tt DIP = 1}, if it is necessary).

\section{Conclusion}

A software model with rather small set of codes providing necessary control flow 
instructions in a reversible processor is considered in presented work. 
Nested procedure calls are implemented using the stack.

\end{document}